\documentstyle[12pt]{article}
\begin{document}
\def\be{\begin{eqnarray}}
\def\en{\end{eqnarray}}
\def\non{\nonumber}
\def\la{\langle}
\def\ra{\rangle}
\def\ep{\varepsilon}
\def\ums{{\mu}_{\rm MS}}
\def\u{\mu_{\rm fact}}
\def\lsim{ {\ \lower-1.2pt\vbox{\hbox{\rlap{$<$}\lower5pt\vbox{\hbox{$\sim$}
}}}\ } }
\def\gsim{ {\ \lower-1.2pt\vbox{\hbox{\rlap{$>$}\lower5pt\vbox{\hbox{$\sim$}
}}}\ } }
\def\getao{g_{_{\eta_0NN}}}
\def\gghost{g_{_{GNN}}}
\def\geta{g_{_{\eta NN}}}
\def\gpi{g_{_{\pi NN}}}
\def\getap{g_{_{\eta'NN}}}
\def\dk{\partial\!\cdot\!K}
\def\pr{{\sl Phys. Rev.}~}
\def\prl{{\sl Phys. Rev. Lett.}~}
\def\pl{{\sl Phys. Lett.}~}
\def\np{{\sl Nucl. Phys.}~}
\def\zp{{\sl Z. Phys.}~}

\font\el=cmbx10 scaled \magstep2
{\obeylines
\hfill IP-ASTP-21-95
\hfill October, 1995}

\vskip 1.5 cm

\centerline{\large\bf U(1) Goldberger-Treiman Relation }
\centerline{\large\bf and Its Connection to the Proton Spin}
\medskip
\bigskip
\medskip
\centerline{\bf Hai-Yang Cheng}
\medskip
\centerline{ Institute of Physics, Academia Sinica}
\centerline{ Taipei, Taiwan 115, Republic of China}
\bigskip
\bigskip
\bigskip
\bigskip
\centerline{\bf Abstract}
\bigskip
{\small

   The U(1) Goldberger-Treiman (GT) relation for the axial charge $g_A^0$ is
reexamined. It is stressed that the isosinglet GT relation in terms of the
$\eta_0$ holds irrespective of the quark masses and the axial anomaly.
We pointed out that the identification of the $\eta_0-N$ and $\dk-N$coupling
terms with the quark and gluon spin components respectively in a proton
is possible but valid only in the chiral-invariant factorization scheme.
In general, the two-component U(1) GT relation can be identified
in a gauge-invariant way with
connected and disconnected insertions. The observation that $(\sqrt{3}f_\pi
/2m_N)\getao$ is related to the connected insertion i.e., the total valence
quark contribution to the proton spin
enables us to determine the physical coupling constants $\getap$
and $\geta$ from the GT relations for $g_A^0$ and $g_A^8$. We found
$\getap=3.4$ and $\geta=4.7\,$.
}

\pagebreak

   {\bf 1}.~~One important thing we learn from the derivation of the
isotriplet Goldberger-Treiman (GT) relation
\be
g_A^3(0)=\,{\sqrt{2}f_\pi\over 2m_N}g_{_{\pi_3NN}},
\en
where $f_\pi=132$ MeV, is that this relation holds irrespective of the
light quark masses. For $m_\pi^2\neq 0$, it is derived through the
use of PCAC; while in the chiral limit, $g_A^3(q^2)$ is related to the
form factor $f_A^3(q^2)q^2$, which receives a nonvanishing pion-pole
contribution even in the $q^2\to 0$ limit. By the same token, it
is tempting to contemplate that the flavor-singlet GT relation
\be
g_A^0(0)=\,{\sqrt{3}f_\pi\over 2m_N}g_{_{\eta_0NN}}^{(0)}
\en
with $\getao^{(0)}$ being a bare direct coupling between $\eta_0$ and the
nucleon, should be also valid irrespective of the meson masses and the axial
anomaly. This is indeed the case: the U(1) GT relation (2) remains totally
unchanged no matter how one varies the anomaly and the quark masses. This
salient feature was first explicitly shown in
[1,2] (see also [3] for a general argument). It was also pointed out
in [4] that this U(1) relation is independent of the interaction of the
ghost field $\partial^\mu K_\mu$ ($K_\mu$ being
the Chern-Simons current) with the nucleon.

Many discussions on the isosinglet GT relation around the period of 1989-1992
[1-7] were mainly motivated by the desire of trying to understand why the
axial charge
$g_A^0$ inferred from the EMC experiment [8] is so small, $g_A^0(0)=0.12\pm
0.18$ at $Q^2=10.7\,{\rm GeV}^2$ (pre-1993).
\footnote{The $q^2$ of the form factor should not be confused with the
momentum transfer $Q^2$ occurred in deep inelastic scttering.}
At first sight, the U(1) GT relation seems not to be in
the right ballpark as the naive SU(6) quark model prediction $g^{(0)}
_{_{\eta_0NN}}=(\sqrt{6}/5)g_{_{\pi NN}}$ yields a too large value of $g_A^0
(0)=0.80\,$. Fortunately, in QCD the ghost field $G\equiv \partial^\mu K_\mu$,
which is necessary for solving the $U_A(1)$ problem, is
allowed to have a direct $U_A(1)$-invariant interaction with the nucleon.
This together with the mixing of $\partial^\mu K_\mu$ with the
$\eta_0$ implies that the net ``physical" $\eta_0-N$ coupling $\getao$ is
composed of the bare coupling $\getao^{(0)}$ and the ghost
coupling $\gghost$. As a consequence, a possible cancellation between
$\getao$ and $\gghost$ terms will render $g_A^0$ smaller. However, this
two-component expression for the axial charge is not free of ambiguity.
For example, $\gghost$ is sometimes assumed to be the
coupling between the glueball and the nucleon in the literature.

    Since the earlier parton-model analysis of polarized deep inelastic
scattering seems to indicate a decomposition of $g_A^0$ in
terms of the quark and gluon spin components [9], this has motivated many
authors to identify the term $(\sqrt{3}f_\pi/2m_N)\getao$ with the total quark
spin $\Delta\Sigma$ in a proton, and the other term with the anomalous
gluon contribution. However, it is also known that the lack of a local
and gauge-invariant operator definition for the quark and gluon spins in
this two-component picture leads to a clash between the OPE approach and
the parton model.  In the former approach, $g_A^0$ is identified with the
total quark spin in a proton. This casts doubt on the usual
two-component interpretation of the axial charge.

   The purpose of this Letter is two-fold. First,
we would like to clarify and present a pertinent physical interpretation for
the two-component isosinglet GT relation. We argue
that the term $(\sqrt{3}f_\pi/2m_N)\getao$ should be identified
with the connected insertion i.e., the total valence quark spin in a
proton. Second, with the valence quark spin inferred from data or from the
quark model, we can employ
the GT relations for $g_A^8$ and $g_A^0$ to determine the physical
coupling constants $\geta$ and $\getap$.

\vskip 0.4cm
{\bf 2}.~~The easist way of deriving the U(1) GT relation is to first work
in the chiral limit. Defining the form factors
\be
\la N(p')|J^5_\mu|N(p)\ra= \,\bar{u}(p')[g_A^0(q^2)\gamma_\mu\gamma_5
+f_A^0(q^2)q_\mu\gamma_5]u(p),
\en
we get
\be
2m_Ng_A^0(0)=\la N|\partial^\mu J^5_\mu|N\ra=3\la N|\dk|N\ra.
\en
Assuming the $\eta_0$ pole dominance for $\dk$, namely
$\dk={1\over\sqrt{3}}m_{\eta_0}^2f_\pi\eta_0$, where the $\eta_0$
mass $m_{\eta_0}$ arises entriely from the axial anomaly, we are led to the
isosinglet GT relation (2). When the quark masses are turned on, chiral
symmetry is explicitly broken but the GT relation in terms of $\eta_0$
remains intact, as shown in [1,2]. Nevertheless, the $\eta_0$ is no longer
a physical meson, and it is related to the mass eigenstates via
\be
\left(\matrix{\pi_3 \cr \eta_8 \cr \eta_0 \cr}\right)=\left(\matrix{ 1 &
\theta_1\cos\theta_3+\theta_2\sin\theta_3 & \theta_1\sin\theta_3-\theta_2
\cos\theta_3 \cr  -\theta_1 & \cos\theta_3 & \sin\theta_3  \cr  \theta_2 &
-\sin\theta_3 & \cos\theta_3 \cr}\right)\left(\matrix{\pi^0 \cr  \eta  \cr
\eta'}\right),
\en
where $\theta_1,~\theta_2$ and $\theta_3$ are the mixing angles of $\pi^0-\eta,
{}~\pi^0-\eta'$ and $\eta-\eta'$ respectively, and their analytic expressions
are given in [6] with the numerical values
\be
\theta_1=-0.016\,,~~~\theta_2=0.0085\,,~~~\theta_3=-18.5^\circ\,.
\en
In Eq.(5) only terms linear in small angles $\theta_1$ and
$\theta_2$ are retained. Consequently, the complete GT relations in terms
of physical coupling constants read [2]
\footnote{For the axial charge $g_A^0$, the authors of [6] obtained a
result something like (see Eq.(24) of the second reference of [6])
\be
{\sqrt{3}f_\pi\over 2m_N}\left({\getap\over \cos\theta_3}-\Delta m_{\eta'}
g_{_{QNN}}\right)-{1\over\sqrt{2}}
g_A^8\tan\theta_3\pm\sqrt{3\over 2}g_A^3(\theta_2-\theta_1\tan\theta_3)
\en
and claimed that in the limit of $\theta_1,~\theta_2\to 0$ but $\theta_3
\neq 0$, it reproduces the result of Veneziano [4] only if the first order
correction from $\theta_3$ (i.e., the $g_A^8\theta_3$ term) is neglected.
However, using Eqs.(8) and (12) one can show that (7) is nothing but
$(\sqrt{3}f_\pi/2m_N)\getao^{(0)}$, as it should be.}
\be
g_A^3(0) ={\sqrt{2}f_\pi\over 2m_N}g_{_{\pi_3NN}} &=& {\sqrt{2}f_\pi\over
2m_N}[\gpi\pm\getap(\theta_1\sin\theta_3-\theta_2\cos\theta_3)  \non \\
&\pm& \geta(\theta_1\cos\theta_3+\theta_2\sin\theta_3)],  \non \\
g_A^8(0)={\sqrt{6}f_\pi\over 2m_N}g_{_{\eta_8NN}} &=& {\sqrt{6}f_\pi\over
2m_N}(\geta\cos\theta_3+\getap\sin\theta_3\mp\gpi\theta_1),   \\
g_A^0(0)={\sqrt{3}f_\pi\over 2m_N}g_{_{\eta_0NN}}^{(0)} &=& {\sqrt{3}f_\pi
\over 2m_N}(\getap\cos\theta_3-\geta\sin\theta_3\pm\gpi\theta_2)+\cdots, \non
\en
where the first sign of $\pm$ or $\mp$ is for the proton and the second
sign for the neutron, and the ellipsis in the GT relation for $g_A^0$ is
related to the ghost coupling, as shown below.
Since the mixing angles $\theta_1$ and $\theta_2$ are
very small, it is evident that
isospin violation in (8) is unobservably small.

   As we have accentuated before, the isosinglet GT relation in terms of the
$\eta_0$ remains unchanged no matter how one varies the quark masses and
the axial anomaly.
\footnote{A smooth extrapolation of the strong coupling constant from on-shell
$q^2$ to $q^2=0$ is understood.}
However, the $\eta_0$ field is subject to a different interpretation
in each different case. For example, when the anomaly
is turned off, the mass of $\eta_0$ is the same as the pion (for
$f_{\eta_0}=f_\pi$). When both quark masses and anomaly are switched off,
the $\eta_0$ becomes a Goldstone boson, and the axial charge at $q^2=0$
receives its contribution from the $\eta_0$ pole.

    When the SU(6) quark model is applied to the coupling $\getao^{(0)}$, it
is evident that the predicted $g_A^0=0.80$ via the GT relation is too large.
This difficulty could be resolved by the observation that {\it a priori}
the ghost field $G\equiv\dk$ is allowed in QCD to have a direct
coupling with the nucleon
\be
{\cal L}=\cdots+{\gghost\over 2m_N}\partial^\mu G\,{\rm Tr}(\bar{N}\gamma_\mu
\gamma_5N)+{\sqrt{3}\over f_\pi}(\dk)\eta_0+\cdots,
\en
so that
\be
\dk=\,{1\over\sqrt{3}}m_{\eta_0}^2f_\pi\eta_0+{1\over 6}\gghost m^2_{\eta_0}
f_\pi\partial^\mu{\rm Tr}(\bar{N}\gamma_\mu\gamma_5N).
\en
However, the matrix element $\la N|\dk|N\ra$ remains unchanged
because of the presence of the $\dk-\eta_0$ mixing, as
schematically shown in Fig.~1:
\be
\la N|\dk|N\ra &=& {1\over\sqrt{3}}f_\pi\getao^{(0)}-{1\over 3}m^2_{\eta_0}
f_\pi\gghost+{1\over 3}m^2_{\eta_0}f_\pi\gghost   \non \\
&=& {1\over\sqrt{3}}f_\pi\getao^{(0)}.
\en
We see that although it is still the bare coupling $\getao^{(0)}$ that
relates to the axial charge $g_A^0$, the ``physical" $\eta_0-N$ coupling
is modified to (see Fig.~1)
\be
\getao=\,\getao^{(0)}+{1\over\sqrt{3}}m_{\eta_0}^2f_\pi\gghost,
\en
where the second term arises from the $\eta_0-\dk$ mixing. As a consequence,
the quark model should be applied to $\getao$ rather than to $\getao^{(0)}$,
and we are led to
\footnote{A two-component expression for the U(1) GT relation was first
put forward by Shore and Veneziano [5].}
\be
g_A^0(0)=\,{\sqrt{3}f_\pi\over 2m_N}(\getao-{1\over \sqrt{3}}m^2_{\eta_0}
f_\pi\gghost).
\en

    It has been proposed that the smallness of $g_A^0$ may be explained
by considering the pole contributions to $\dk$ from higher single particle
states $X$ above the $\eta_0$, so that the isosinglet GT relation has the
form (see e.g., Chao {\it et al.} [4], Ji [4], Bartelski and Tatur [10])
\be
g_A^0(0)=\,{\sqrt{3}\over 2m_N}(f_{\eta_0}\getao+\sum_X f_Xg_{_{XNN}}).
\en
The state $X$ could be the radial excitation state of $\eta_0$ or a $0^{-+}$
glueball. (Note that the ghost field $\dk$ is {\it not} a physical glueball
as it can be eliminated via the equation of motion.)
However, we will not pursue this possibility further for two reasons: (i)
It is entirely unknown whether or not the $X$ states contribute destructively
to $g_A^0$. (ii) As we shall see later, the contribution from a direct
interaction of the ghost field with the nucleon corresponds to a
disconnected insertion, which is shown to be negative according to
recent lattice QCD calculations [11,12]. Therefore, the ghost-field effect
is realistic, and if the contributions due
to the states $X$ are taken into account, one should make the following
replacement
\be
\getao\to\getao-{1\over\sqrt{3}}m^2_{\eta_0}f_\pi\gghost,~~~~g_{_{XNN}}\to
g_{_{XNN}}-{1\over \lambda}m_X^2g_{_{XNN}}
\en
in Eq.(14), where $\lambda$ is the $\dk\!-\!X$ mixing.

\vskip 0.4cm
{\bf 3}.~~It has been claimed in the parton-model study of
polarized deep inleastic scattering
that $g_A^0$ is related to the flavor-singlet quark spin and the
anomalous gluon contribution [9]:
\be
g_A^0(0)=\Delta u'+\Delta d'+\Delta s'-{3\alpha_s\over 2\pi}\Delta G
\equiv \Delta\Sigma'-\Delta\Gamma,
\en
where $\Delta q'=q^\uparrow+\bar{q}^\uparrow-q^\downarrow-\bar{q}^\downarrow$
is the net helicity of the quark flavor $q$ in a proton,
and $\Delta G=G^\uparrow-G^\downarrow$. In Eq.(16) a superscript ``prime"
is used to denote a quark spin different from the one appearing in the
OPE approach (see below). By comparing (16) with (13),
it is tempting to identify the two components of the U(1) GT relation as
\be
\Delta\Sigma'={\sqrt{3}f_\pi\over 2m_N}\getao,~~~~\Delta\Gamma={m_{\eta_0}
^2f_\pi^2\over 2m_N}\gghost.
\en
On the contrary, in the OPE approach only the quark operator contributes to
the first moment of the proton structure function $g^p_1(x)$ at the twist-2
and spin-1 level [13], so that
\be
g_A^0(0)=\Delta u+\Delta d+\Delta s\equiv\Delta\Sigma.
\en
Therefore, one may wonder if the identification (17) is unique and sensible.

    The above issue has to do with whether or not gluons contribute to
$\Gamma_1^p$, the first moment of the polarized proton structure function
$g_1^p(x)$.
Since this issue has been addressed and resolved by Bodwin and Qiu [14],
in the following we will simply outline the main arguments (see also [15]).

    The gluonic contribution to $\Gamma_1^p$ is governed by the first moment
of the differential polarized photon-gluon scattering cross section denoted
by $\Delta\sigma(x)$. A direct calculation of the photon-gluon scattering
box diagram shows that $\Delta\sigma
(x)$ has collinear and infrared singularities at $m^2=p^2=0$, with $m$ the
quark mass and $p$ the momentum of the gluon. With two different choices
of the soft cutoff, one obtains
\be
\Delta\sigma_{\rm CCM}(x)=\,(1-2x)\left(\ln{Q^2\over -p^2}+\ln{1\over x^2}-2
\right),
\en
for $m^2=0$ and $p^2\neq 0$ (Carlitz {\it et al.} [9]), and
\be
\Delta\sigma_{\rm AR}(x)=\,(1-2x)\left(\ln{Q^2\over m^2}+\ln{1-x\over x}-1
\right)-2(1-x),
\en
for $m^2\neq 0$ and $p^2=0$ (Altarelli and Ross [9]).
At first sight, it appears that $\int^1_0\Delta\sigma^{\rm hard}(x)dx=1$
in both regulator schemes because the $2(1-x)$ term in
$\Delta\sigma_{\rm AR}(x)$ arising from $k_\perp^2\sim m^2$ is a soft
contribution and because the $\ln(Q^2/\!-\!p^2)$ and $\ln(Q^2/m^2)$ terms,
which depend logarithmically on the soft cutoff, make no contribution to
the first momemt due to chiral symmetry or helicity conservation, recalling
the splitting function $\Delta P_{qG}(x)={1\over 2}(2x-1)$.
However, the cancellation of the soft contribution from different $x$ regions
is not reliable because chiral symmetry may be broken at some hadronic scale
$\Lambda$ through some nonperturbative effects. As a consequence, one
has to introduce a factorization scale $\u$ to subtract the unwanted soft
contribution, i.e., the contribution arising from the distribution of quarks
and antiquarks in a gluon:
\be
\Delta\sigma^{\rm hard}(x,Q^2/\u^2)=\,\Delta\sigma(x,Q^2)-\Delta\sigma^{\rm
soft}(x,\u^2).
\en

In practice, one makes an approximate expression for the box diagram that is
valid for $k_\perp^2<<Q^2$ and then introduces an ultraviolet cutoff on the
integration variable $k_\perp$
to ensure that only the region $k^2_\perp\lsim\u^2$ contributes to the soft
part [14]. The choice of the regulator specifies the factorization convention.
There are two sources contributing to the first moment of $\Delta\sigma(x)$:
one from $k_\perp^2\sim Q^2$ and the other from chiral symmetry breaking.
When the ultraviolet cutoff is gauge invariant, it breaks chiral symmetry
due to the presence of the axial anomaly and hence
makes a contribution to $\Delta\sigma^{\rm soft}$. So we have
\be
\int^1_0\Delta\sigma_{\rm CCM}^{\rm soft}(x)dx=1,~~~~~\int^1_0\Delta\sigma_{
\rm AR}^{\rm soft}(x)dx=0.
\en
In the mass-regulator scheme, the original soft contribution coming from
$k^2_\perp\sim m^2$, where chiral symmetry is explicitly broken by the
quark mass, is canceled by the contribution arising from chiral symmetry
breaking induced by the ultraviolet cutoff. Therefore, in the gauge-invariant
factorization scheme
$\int^1_0\Delta\sigma^{\rm hard}(x)dx=0$ and hence $g_A^0(0)=\Delta\Sigma$.
In this scheme, the quark spin has a gauge-invariant local operator
definition: $s_\mu\Delta q=\la p|\bar{q}\gamma_\mu\gamma_5q|p\ra$.
It is $Q^2$ dependent because of the nonvanishing anomalous
dimension associated with the flavor-singlet quark operator.
By contrast, it is also possible to choose a
chiral-invariant but gauge-variant ultraviolet cutoff, so that
\be
\int^1_0\Delta\tilde{\sigma}_{\rm CCM}^{\rm soft}(x)dx=0,~~~~~\int^1_0\Delta
\tilde{\sigma}_{\rm AR}^{\rm soft}(x)dx=1.
\en
This together with Eqs.(19) and (20) leads to
$\int^1_0\Delta\tilde{\sigma}^{\rm hard}(x)dx=1$. It is thus evident
that gluons contribute to $\Gamma_1^p$ and
$g_A^0(0)=\Delta\Sigma'-\Delta\Gamma$ in the chiral-invariant
factorization scheme. Contrary to the first scheme, $\Delta q'$ here cannot
be written as a matrix element of a gauge-invariant local operator; it is
either gauge variant or involves a nonlocal operator. Moreover, $\Delta q'$
is $Q^2$ independent as the gauge-variant ultraviolet cutoff in this
scheme does not flip helicity. It is thus close to the naive
intuition in the parton model that the
quark helicity is not affected by gluon emissions.

   It is clear that the issue of whether or not
gluons contribute to $\Gamma_1^p$ is purely a matter of the factorization
scheme chosen in defining the quark spin density
\footnote{In principle, the choice of $\Delta q$ and $\Delta\sigma^{\rm hard}
(x)$ or $\Delta q'$ and $\Delta\tilde{\sigma}^{\rm hard}(x)$ is just a matter
of convention. In practice, the gauge-invariant $\Delta q$ is probably more
useful than the $Q^2$-independent $\Delta q'$ since the former can be
expressed as a nucleon matrix element of a local gauge-invariant operator and
is thus calculable in lattice QCD. Moreover, the polarized Altarelli-Parisi
equations cannot be applied to $\Delta q'$ directly [16]. It has been
advocated that $\Delta q'$ and $\Delta G$ have a simple partonic definition:
the former (latter) can be identified in one-jet (two-jet) events in
polarized deep inelastic scattering (Carlitz {\it et al.} [9]). However,
as pointed out in [17], it is impossible to separate the jets when the target
is at rest because the longitudinal momentum is of order $Q^2/M$, whereas
the transverse momentum $k_\perp$ can only be of order $Q$. Consequently,
the $q$ and $\bar{q}$ jets are collinear even they may have large transverse
momentum.}
and the hard gluon-photon scattering cross section [14]. We thus conclude that
the identification of the U(1) GT relation with the quark and gluon
spin components in a proton as given in Eq.(17) is possible but valid
only in the chiral-invariant factorization scheme.
Next, one may ask what will be the physical interpretation for the
gauge-invariant $\getao$ and $\gghost$ terms in the two-component isosinglet
GT relation (13) in the gauge-invariant
factorization scheme in which $g_A^0=\Delta\Sigma$ ? We note that the
evaluation of the hadronic flavor-singlet current involves a disconnected
insertion in addition to the connected one (see Fig.~2). The connected and
disconnected insertions are related to valence quark and vacuum polarization
(i.e., sea quark) contributions respectively (Liu [4]) and are separately
gauge invariant. A recent lattice calculation [10] shows a sea
polarization in a polarized proton: $\Delta u_s=
\Delta d_s=\Delta s=-0.12\pm 0.01$ from the disconnected contribution.
This empirical SU(3)-flavor symmetry for sea polarization, which is known
to be not true for the unpolarized counterpart,
implies that {\it the disconnected insertion is dominated by the
axial anomaly of the triangle diagram}. Since the triangle contribution is
proportional to $\dk$, the ghost field, it is thus quite natural to make
the gauge invariant identification:
\be
{\sqrt{3}f_\pi\over 2m_N}\getao=\,{\rm connected~insertion},~~~~-{m_{\eta_0}
^2f_\pi^2\over 2m_N}\gghost=\,{\rm disconnected~insertion},
\en
which is valid in both factorization schemes. In the gauge-invariant
factorization scheme, the disconnected insertion, which is responsible for
the smallness of $g_A^0$, should be interpreted as a screening effect for
the axial charge owing to the negative sea polarization rather than an
anomalous gluonic effect.

\vskip 0.4cm
{\bf 4.}~~Having identified the two-component U(1) GT relation (13) with
connected and disconnected insertions, we are now able to extract the
physical coupling constants $\getap$ and $\geta$. This is because the
connected insertion (CI) corresponds to the total valence quark contribution
to the proton spin, so it is related to the quark model expectation;
that is,
\be
{\sqrt{3}f_\pi\over 2m_N}\getao=g_A^0({\rm CI})=\,\Delta u_v+\Delta d_v=\,
3F-D,
\en
where last identity follows from the fact that in the quark model $g_A^8=3F-D
=\Delta u_v+\Delta d_v$. Another way to see this is that $g_A^8=
\Delta u+\Delta d-2\Delta s\to\Delta u_v+\Delta d_v$ due to the aforementioned
SU(3) symmetry for sea polarization. Unlike the previous identification (17),
$g_A^0({\rm CI})$ here is {\it not} identified with the total quark spin
$\Delta\Sigma$. In the nonrelativistic quark limit, $F={2\over 3},~D=1$,
and hence $\Delta u_v+
\Delta d_v=1$. With the inclusion of the relativistic effects,
$F$ and $D$ are reduced to $F=0.459$ and $D=0.798$
without including errors [18], and $g_A^0({\rm CI})$ is reduced to a value
of $0.579\,$.

   From Eqs.(8) and (25), the GT relations for $g_A^8$ and $g_A^0$ are recast
to
\be
&& 3F-D={\sqrt{6}f_\pi\over 2m_N}g_{_{\eta_8NN}}= {\sqrt{6}f_\pi\over
2m_N}(\geta\cos\theta_3+\getap\sin\theta_3),   \non \\
&& 3F-D={\sqrt{3}f_\pi\over 2m_N}g_{_{\eta_0NN}} = {\sqrt{3}f_\pi
\over 2m_N}(\getap\cos\theta_3-\geta\sin\theta_3),
\en
where the tiny isospin-violating effect has been neglected.
Note that we have $\getao$ instead of $\getao^{(0)}$ on the second line
of the above equation. Using $\theta_3=-18.5^\circ$ [see Eq.(6)], it
follows from (26) that
\be
\getap=3.4\,,~~~~~\geta=4.7\,,
\en
while
\be
\getao=4.8\,,~~~~~g_{_{\eta_8NN}}=3.4\,.
\en
It is interesting to note that we have $\getap<\geta$, whereas $\getao>
g_{_{\eta_8NN}}$.
Phenomenologically, the determination of $\getap$ and $\geta$ is rather
difficult and subject to large uncertainties. The analysis of the $NN$
potential yields $\getap=7.3$ and $\geta=6.8\,$ [19], while the forward
$NN$ scattering analyzed using dispersion relations gives $\getap,~\geta
<3.5$\, [20]. But these analyses did not take into account the ghost
pole contribution. An estimate of the $\eta'\to 2\gamma$ decay rate through
the baryon triangle contributions yields $\getap=6.3\pm 0.4$ [21].

   Finally, the ghost coupling is determined from the disconnected insertion
(DI)
\be
-{m_{\eta_0}^2f_\pi^2\over 2m_N}\gghost=\,g_A^0({\rm DI})
=\,\Delta u_s+\Delta d_s+\Delta s\to 3\Delta s.
\en
A combination of all EMC, SMC, E142 and E143 data for $\Gamma_1^p$ at $\la
Q^2\ra=10\,{\rm GeV}^2$ yields [22]
\be
\Delta u=\,0.83\pm 0.02\,,~~~\Delta d=-0.43\pm 0.02\,,~~~\Delta s=-0.09\pm
0.02\,,
\en
and hence
\be
g_A^0(0)=\Delta\Sigma=\,0.31\pm 0.06\,.
\en
{}From (29) and (30) we obtain
\be
\gghost\approx 55\,{\rm GeV}^{-3}.
\en

\vskip 0.4cm
{\bf 5.}~~To summarize, we have emphasized that the U(1) GT relation in
terms of the $\eta_0$ remains totally unchanged no matter how one varies
the quark masses and the axial anomaly, and pointed out that the
two-component expression of the isosinglet GT relation should be identified
with the connected and disconnected insertions; the identification with the
quark and gluon spin components in a proton is possible only in the
chiral-invariant
factorization scheme. Since $(\sqrt{3}f_\pi/2m_N)\getao$ is related to the
total valence quark contribution to the proton spin, we have determined the
physical coupling constants $\getap$ and $\geta$ from the GT relations for
$g_A^0$ and $g_A^8$ and found that $\getap=3.4$ and $\geta=4.7\,$.

\vskip 1.8cm
\centerline{\bf ACKNOWLEDGMENT}
\vskip 0.3 cm
    This work was supported in part by the National Science Council of ROC
under Contract No. NSC84-2112-M-001-014.

\vskip 1.5 cm
\centerline{\bf REFERENCES}
\vskip 0.3 cm
\begin{enumerate}

\item J. Schechter, V. Soni, A. Subbaraman, and H. Weigel, \prl {\bf 65}, 2955
(1990); {\sl Mod. Phys. Lett.} {\bf A5}, 2543 (1990);  {\sl Mod. Phys.
Lett.} {\bf A7}, 1 (1992).

\item J. Bartelski and S. Tatur, \pl {\bf B265}, 192 (1991).

\item G.M. Shore and G. Veneziano, \np {\bf B381}, 23 (1992).

\item G. Veneziano, {\sl Mod. Phys. Lett.} {\bf A4}, 1605 (1989); T.D. Cohen
and M.K. Banerjee, \pl {\bf B230}, 129 (1989); T. Hatsuda,
\np {\bf B329}, 376 (1990); X. Ji, \prl {\bf 65}, 408 (1990); M. Birse, \pl
{\bf B249}, 291 (1990); K.T. Chao, J. Wen, and H. Zeng, \pr {\bf D46},
5078 (1992); M. Wakamatsu, \pl {\bf
B280}, 97 (1992); K.F. Liu, \pl {\bf B281}, 141 (1992).

\item G.M. Shore and G. Veneziano, \pl {\bf B244}, 75 (1990).

\item A.V. Efremov, J. Soffer, and N.A. T\"ornqvist, \prl {\bf 64}, 1495
(1990); \pr {\bf D44}, 1369 (1991).

\item T. Hatsuda, {\sl Nucl. Phys.~(Proc. Suppl.)} {\bf 23B}, 108 (1991).

\item EMC Collaboration, J. Ashman {\it et al.,} \np {\bf B238}, 1 (1990); \pl
{\bf B206}, 364 (1988).

\item G. Altarelli and G.G. Ross, \pl {\bf B212}, 391 (1988); R.D. Carlitz,
J.C. Collins, and A.H. Mueller, \pl {\bf B214}, 229 (1988);
A.V. Efremov and O.V. Teryaev, in {\it Proceedings of the International
Hadron Symposium}, Bechyne, Czechoslovakia, 1988, eds. Fischer {\it et al.}
(Czechoslovakian Academy of Science, Prague, 1989), p.302.

\item J. Bartelski and S. Tatur, \pl {\bf B305}, 281 (1993).

\item S.J. Dong, J.-F. Laga\"e, and K.F. Liu, \prl {\bf 75}, 2096 (1995).

\item M. Fukugita, Y. Kuramashi, M. Okawa, and A. Ukawa, \prl {\bf 75}, 2092
(1995).

\item R.L. Jaffe and A.V. Manohar, \np {\bf B337} 509 (1990).

\item G.T. Bodwin and J. Qiu, \pr {\bf D41}, 2755 (1990), and
in {\it Proc. Polarized Collider Workshop}, University Park, PA, 1990, eds.
J. Collins {\it et al.} (AIP, New York, 1991), p.285.

\item H.Y. Cheng, H.H. Liu, and C.Y. Wu, IP-ASTP-17-95 (1995).

\item S.D. Bass and A.W. Thomas, {\sl J. Phys.} {\bf G19}, 925 (1993);
Cavendish preprint 93/4 (1993).

\item A.V. Manohar, \pl {\bf B255}, 579 (1991).

\item F. Close and R.G. Roberts, \pl {\bf B316}, 165 (1993).

\item O. Dumbrajs {\it et al.,} \np {\bf B216}, 277 (1983).

\item W. Brein and P. Knoll, \np {\bf A338}, 332 (1980).

\item B. Bagchi and A. Lahiri, {\sl J. Phys.} {\bf G16}, L239 (1990).

\item C.Y. Prescott, SLAC-PUB-6620 (1994); J. Ellis and M. Karliner,
\pl {\bf B341}, 397 (1995).

\end{enumerate}
\vskip 1.3 cm
\centerline{\bf FIGURE CAPTIONS}
\vskip 0.3cm
\begin{description}

\item[Fig.~1] Contributions to the matrix element $\la N|\dk|N\ra$ from (1)
the $\eta_0$ pole dominance, (2) a direct coupling of the ghost field
with the nucleon, and (3) the $\dk-\eta_0$ mixing.

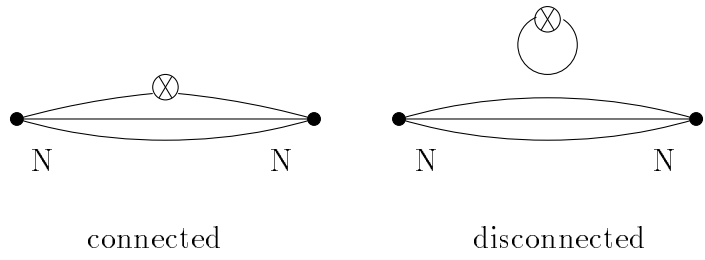
\item[Fig.~2] Connected and disconnected insertions.

\end{description}

\end{document}